\documentclass[pra,aps,twocolumn,floatfix,showpacs,tightenlines,amsmath,amssymb,superscriptaddress]{revtex4}

\usepackage{amsmath}
\usepackage{epsfig}

\usepackage{graphicx}% Include figure files
\usepackage{dcolumn}% Align table columns on decimal point
\usepackage{bm}% bold math

\newcommand{\ket}[1]{|{#1}\rangle}
\newcommand{\bra}[1]{\langle{#1}|}

\begin{document}

\title{A naturally error suppressing quantum memory}

\author{Fumiko Yamaguchi}\email{yamaguchi@stanford.edu}
\affiliation{E. L. Ginzton Laboratory, Stanford University,
Stanford, CA 94305, USA}
\author{Yoshihisa Yamamoto}
\affiliation{E. L. Ginzton Laboratory, Stanford University,
Stanford, CA 94305, USA} \affiliation{National Institute of
Informatics, Hitotsubashi, Chiyoda-ku, Tokyo 101-8430, Japan}

\date{\today}

\begin{abstract}
We propose a method to construct quantum storage wherein the phase
error due to decoherence is naturally suppressed without constant
error detection and correction. As an example, we describe a
quantum memory made of two physical qubits encoded in the ground
state of a two-qubit phase-error detecting code. Such a system can
be simulated by introducing a coupling between the two physical
qubits. This method is effective for physical systems in which the
$T_1$ decay process is negligible but coherence is limited by the
$T_2$ decay process. We take trapped ions as a possible example to
apply the natural suppression method and show that the $T_2$ decay
time due to slow ambient fluctuating fields at the physical qubits
can be lengthened as much as $10^4$.
\end{abstract}


\maketitle

%\section{Introduction}

Decoherence in quantum bits (qubits), even if causing only a
single qubit error, collapses an exponentially large amount of
data and jeopardizes the potential power of quantum computation
\cite{Grover1997,Shor1994}. Such decoherence also limits the
long-lived quantum storage necessary for long-distance quantum
communication and distributed quantum computation
\cite{Bennett1984,Grover1997B,Briegel1998,Bennett1996}.

Quantum error correction
\cite{Shor1995,Steane1996B,Calderbank1996,Steane1996} may allow us
to regain information from a collapsed state even in the presence
of such errors. By fault-tolerant quantum computation, the concept
introduced in \cite{Shor1996}, arbitrarily long quantum
computation can be performed even with imperfect logic gates,
under the assumption that the error per quantum gate and per qubit
during a logic gate is below a threshold value
\cite{Kitaev1997,Preskill1998,Knill1998,Aharonov1998,Kitaev1997B}.
However, the resource overhead for fault-tolerant error correction
is likely to be impractical to implement in an actual physical
system.

Storing quantum information in a decoherence-free subspace (DFS)
\cite{Zanardi1997,Lidar1998,Duan1998,Nielsen2000} is an excellent
way to reduce this overhead. Instead of active error detection and
correction using numerous ancilla qubits, which would have been
fatal to a conventional fault-tolerant error correction method,
the technique uses symmetry of the system such that in the
subspace certain causes of decoherence disappear.

One alternative method is natural error suppression by energy
consideration \cite{Kitaev1997B,Barnes2000,Bacon2001}. The method
incorporates qubit states encoded so that any error that collapses
an encoded qubit state costs energy. Unless that amount of energy
is supplied by the environment, such an error is suppressed. As in
the DFS method, this natural error suppression method eliminates
the need for frequent measurement and logic operation to correct
quantum errors. This also improves a possible weakness of the DFS
method -- the coupling of qubits to the environment does not need
to possess a certain symmetry required by the DFS method. In
addition, if we design a physical system whose ground state is the
code subspace of a well-studied error correcting code, for which
fault-tolerant logic operations are known, the logic operations on
the encoded qubits are exactly the same as the ones for the error
correcting code.

Natural error suppression and correction have been discussed in
the context of errors caused by a thermal reservoir
\cite{Barnes2000,Bacon2001}, and therefore the condition derived
for the method to work is that the energy cost of an error must be
greater than the thermal energy. However, in many physical
systems, this condition is difficult to satisfy and thus natural
error suppression may be ineffective. In this paper, we propose a
natural error suppression method to reduce decoherence resulting
from low-frequency field-fluctuation noise. The condition for this
natural error suppression is derived by comparison of the energy
cost and the cut-off frequency of the field-fluctuation noise.
Even a small energy cost would decrease the effect of slow
field-fluctuation noise. Trapped ions serve as a good example of
the effectiveness of the method, since the main source of
decoherence in quantum memories is slow ambient magnetic-field
fluctuation, and any decoherence due to coupling to a thermal
reservoir is negligible
\cite{Fisk1995,Wineland1998,Kielpinski2002,
Rowe2002,Schmidt-Kaler2003,Wineland2003,Chiaverini2005}. In such
systems, the $T_1$ decay time is extremely long and bit-flip
errors are unlikely to occur. Therefore, we consider constructing
an error suppressing memory based on a two-qubit phase-flip error
detecting code, and illustrate how much the $T_2$ decay time can
be lengthened by this method. This will provide a method to design
quantum memories with reduced decoherence, instead of -- or in
addition to -- other methods such as applying spin echo technique
\cite{Andersen2003,Riebe2004,Barrett2004} and using
magnetic-field-independent transitions \cite{Langer2005}.

%\section{Natural error suppression}
%\subsection{Two-qubit phase-flip detecting code}
When a logical qubit is encoded as $\ket{0}_{L}=\ket{00}$ and
$\ket{1}_{L}=\ket{11}$ using two physical qubits, at most one
bit-flip error ($\sigma_{1x}$, or $\sigma_{2x}$) can be detected.
Here we denote our physical qubit basis states as $\ket{0}$ and
$\ket{1}$, and the associated Pauli matrices as $\sigma_{iq}$
($q=x,y, z$) for qubit $i$ ($i=1, 2$). The code subspace
$\{\ket{0}_{L}, \ket{1}_{L}\}$ is a set of simultaneous
eigenstates of the stabilizer generator $g=\sigma_{1z}\sigma_{2z}$
with eigenvalue $+1$. A bit-flip error removes the encoded state
from the code subspace, and the stabilizer generator has
eigenvalue $-1$.
%($\ket{01}$ or $\ket{10}$)
%
This code can be converted into a phase-flip detecting code
(detecting $\sigma_{1z}$, or $\sigma_{2z}$) by encoding a logical
qubit as $\ket{0}_{L}=\ket{++}$ and $\ket{1}_{L}=\ket{--}$ in the
rotated basis states, $\ket{\pm}=(\ket{0}\pm\ket{1})/\sqrt{2}$.
The stabilizer generator for this code is
$g=\sigma_{1x}\sigma_{2x}$. We use this two-qubit phase-flip
detecting code to design a naturally error suppressing quantum
memory.

%\subsection{Error suppression Hamiltonian}
Associated with the two-qubit phase-flip detecting code, we
consider a physical system with an interaction between qubits 1
and 2 in the form of
\begin{equation}
    H_{\rm ES} = - 2J I_{1x}I_{2x},
    \label{eq:error_suppression_hamiltonian}
\end{equation}
where the qubits are two spins $I_{1x}=\frac{1}{2}\sigma_{1x}$,
$I_{2x}=\frac{1}{2}\sigma_{2x}$. The ground states comprise the
code subspace of the two-qubit phase-flip error detecting code,
and excited states correspond to states collapsed from the encoded
state due to a phase-flip error. Therefore, unless the energy
difference $J$ is supplied from the environment, a phase-flip
error is suppressed automatically without the need for
measurements to detect the error or logic operations to correct
it.

%\subsection{Suppression of errors due to decoherence}
The system under consideration consists of trapped ions, the two
hyperfine ground states of which are the physical qubit states. A
quantum memory is composed of two such qubits, described by the
spin 1/2 operators,
\begin{equation}
    H_0 = \omega_0 (I_{1z} + I_{2z}),
\end{equation}
where $\omega_0$ is the hyperfine splitting, and interaction of
the qubits with the fluctuating field $H_{iq}(t)$ at qubit $i$
($i=1,2$) in the $q$-direction ($q=x,y,z$),
\begin{equation}
    H_1(t) = \gamma \sum_{i,q}H_{iq}(t)I_{iq},
\end{equation}
where $\gamma$ is the gyro-magnetic ratio, in addition to the
error suppression Hamiltonian
(\ref{eq:error_suppression_hamiltonian}). Here we assume that the
hyperfine splitting term $H_0$ is much larger than the other
terms, $H_{\rm ES}$ and $H_1(t)$.  In ion trap experiments, the
hyperfine splitting is about 10GHz and the amplitude of magnetic
field fluctuation is about 2kHz
\cite{Wineland1998,Schmidt-Kaler2003}. Then the equation of motion
of the density matrix \cite{Slichter1996}
\begin{equation}
    \frac{{\rm d}\rho(t)}{{\rm d}t} = i[\rho(t), H_0 + H_1(t) + H_{\rm
    ES}]
\end{equation}
will lead to the equation of motion in the interaction picture in
$H_0$, denoted with $\ast$,
\begin{equation}
   \frac{{\rm d}\rho^{\ast}(t)}{{\rm d}t} = i[\rho^{\ast}(t), H^{\ast}_1(t) + H^{\ast}_{\rm
    ES}(t)],
\end{equation}
where $\rho^{\ast}(t) = e^{iH_0t}\rho(t) e^{-iH_0t}$,
$H_1^{\ast}(t) = e^{iH_0t}H_1(t) e^{-iH_0t}$ and so on. We further
assume that the fluctuating field amplitude is smaller than the
interaction between the two qubits and obtain the second-order
perturbation expansion,
\begin{equation}
    \frac{{\rm d}\tilde{\rho}^{\ast}(t)}{{\rm d}t} = i[\tilde{\rho}^{\ast}(0), \tilde{H}^{\ast}_1(t)]
    - \int_{0}^{t}[[\tilde{\rho}^{\ast}(0),
    \tilde{H}_1^{\ast}(t-\tau)],\tilde{H}_1^{\ast}(t)]{\rm d}\tau,
    \label{eq:perturbation}
\end{equation}
where $\tilde{\rho}^{\ast} = e^{iH_{\rm
ES}^{\ast}t}\rho^{\ast}e^{-iH_{\rm ES}^{\ast}t}$=
$e^{iH_0t}e^{iH_{\rm ES}t}\rho e^{-iH_{\rm ES}t}e^{-iH_0t}$ and
$\tilde{H}_1^{\ast}(t) = e^{iH_{\rm
ES}^{\ast}t}H_1^{\ast}(t)e^{-iH_{\rm ES}^{\ast}t}$.

To compute the second term in the right-hand side of
Eq.~(\ref{eq:perturbation}), we note that
\begin{equation}
    \tilde{H}_1^{\ast}(t) = e^{iH_0t}e^{iH_{\rm
    ES}t}H_1(t)e^{-iH_{\rm ES}t}e^{-iH_0t},
\end{equation}
and
\begin{eqnarray}
    e^{iH_{\rm ES}t}H_1(t)e^{-iH_{\rm ES}t} = \gamma
    \sum_{iq}H_{iq}K_{iq}(t),
\end{eqnarray}
where
\begin{eqnarray}
    && K_{ix}(t) = I_{ix},\nonumber\\
    &&K_{iy}(t) = I_{iy}\cos Jt + 2I_{iz}I_{\bar{i}x}\sin Jt,\\
    &&K_{iz}(t) = I_{iz}\cos Jt - 2I_{iy}I_{\bar{i}x}\sin Jt.
    \nonumber
\end{eqnarray}
Here $\bar{i}=2$ and 1 for $i=1$ and 2, respectively. The equation
of motion of a matrix element $\tilde{\rho}_{\alpha\alpha'}$ is
then
\begin{widetext}
\begin{eqnarray}
    &&\frac{{\rm d}\tilde{\rho}^{\ast}_{\alpha\alpha'}(t)}{{\rm d}t}
    = \gamma^2\sum_{\beta,\beta',i,j,q}\int_0^t{\rm d}\tau e^{i(E_{\alpha}-E_{\beta}+E_{\beta'}-E_{\alpha'})t}
    \bra{\alpha}K_{iq}(t-\tau)\ket{\beta}\tilde{\rho}^{\ast}_{\beta\beta'}(0)
    \bra{\beta'}K_{jq}(t)\ket{\alpha'}\nonumber\\
    &&\quad\times[e^{-i(E_{\alpha}-E_{\beta})\tau}+e^{-i(E_{\beta'}-E_{\alpha'})\tau}]
    \overline{H_{iq}(t-\tau)H_{jq}(t)}\nonumber\\
    &&-\gamma^2\sum_{\beta,\beta',i,j,q}\int_0^t{\rm d}\tau
    [e^{i(E_{\beta}-E_{\alpha'})t}\tilde{\rho}^{\ast}_{\alpha\beta}(0)\bra{\beta}K_{iq}(t-\tau)\ket{\beta'}
    \bra{\beta'}K_{jq}(t)\ket{\alpha'}\nonumber\\
   && + e^{i(E_{\alpha}-E_{\beta'})t} \bra{\alpha}K_{iq}(t)\ket{\beta}\bra{\beta}K_{jq}(t-\tau)\ket{\beta'}
    \tilde{\rho}^{\ast}_{\beta'\alpha'}(0)]%\nonumber\\
    %&&\quad\times
    e^{-i(E_{\beta}-E_{\beta'})\tau}\overline{H_{iq}(t-\tau)H_{jq}(t)},
    \label{eq:perturbation2}
\end{eqnarray}
\end{widetext}
where $\ket{\alpha}$ ($\alpha=\pm 1/2$) is an eigenstate of $H_0$
satisfying $H_0\ket{\alpha}=E_{\alpha}\ket{\alpha}$
($E_{\alpha}=\omega_0\alpha$). Furthermore, we took the time
average of fluctuating field and assumed the time average of the
fluctuating field is zero
\begin{equation}
    \overline{H_{iq}(t)}= 0,
\end{equation}
and the fluctuations of the three components of the field are
independent,
\begin{equation}
    \overline{H_{iq}(t-\tau)H_{jq'}(t)}=0, \quad\mbox{for }
    q\neq q'.
\end{equation}
We further simplify Eq.~(\ref{eq:perturbation2}) by ignoring terms
oscillating at $\pm\omega_0 t$ or faster, and assuming the
correlation time of $\tau_0$ of the spectral densities of the
fluctuating fields are longer than $1/J$. The assumption allows us
to drop the terms oscillating at $\pm 2Jt$, and then we find the
equation of motion for $\langle I_{ix}\rangle = {\rm Tr}(\rho
I_{ix})$,
\begin{eqnarray}
    \lefteqn{\frac{{\rm d}\langle I_{ix}\rangle}{{\rm d}t}
    = i\sum_{\alpha,\alpha'}[\rho, H_0 + H_{\rm
    ES}]_{\alpha,\alpha'}
    \bra{\alpha'}I_{ix}\ket{\alpha}} \nonumber\\
    &&-\langle I_{ix}\rangle
    \gamma^2\left[\frac{k_{yy}(\omega_0+J)+k_{yy}(\omega_0-J)}{2}+k_{zz}(J)\right],
    \nonumber\\
\end{eqnarray}
where the spectral densities of the fluctuating fields are defined
as
\begin{equation}
    k_{qq}^{ij}(\omega)=\frac{1}{2}\int_{-\infty}^{+\infty}
    \overline{H_{iq}(t-\tau)H_{jq}(t)}e^{-i\omega\tau}{\rm d}\tau.
\end{equation}
We have assumed the time average is independent of $t$ and zero
when $\tau$ is greater than a certain critical value. Therefore,
decay time of $\langle{I_{1x}}\rangle$ is
\begin{equation}
    \frac{1}{T_{2x}}=
    \gamma^2\left[\frac{k_{yy}(\omega_0+J)+k_{yy}(\omega_0-J)}{2}+k_{zz}(J)\right].
    \label{eq:T2}
\end{equation}
Note that only the fluctuating field component at qubit 1
$k^{11}_{yy}$ remains in Eq.~(\ref{eq:T2}). In the following, we
assume that $k^{11}_{qq} = k^{22}_{qq}(\omega) \equiv
k_{qq}(\omega)$ for simplicity. Similarly, decay time $T_{2y}$ for
$\langle{I_{1y}}\rangle$ and $T_1$ for $\langle I_{1z}\rangle$ are
obtained as
\begin{eqnarray}
    \frac{1}{T_{2y}}&=&
    \gamma^2\left[k_{xx}(\omega_0)+k_{zz}(J)\right],\\
    \frac{1}{T_{1}}&=&
    \gamma^2\left[
    k_{xx}(\omega_0)+\frac{k_{yy}(\omega_0+J)+k_{yy}(\omega_0-J)}{2}\right].
    \nonumber\\
\end{eqnarray}
The $T_2$ process averaged over procession in the transverse
direction is
\begin{equation}
    \frac{1}{T_2} = \frac{1}{2}\left(
        \frac{1}{T_{2x}}+\frac{1}{T_{2y}}
    \right) = \frac{1}{2T_1}+\gamma^2 k_{zz}(J).
\end{equation}
These results should be compared to the decay times $T_1^{0}$ and
$T_{2}^0$ in the case where there is no error suppression
Hamiltonian,
\begin{equation}
    \frac{1}{T_{1}^0}=
    \gamma^2\left[k_{xx}(\omega_0)+k_{yy}(\omega_0)\right],
    \quad
    \frac{1}{T_2^0} = \frac{1}{2T_1^0}+\gamma^2k_{zz}(0).
\end{equation}

We now assume a simple exponential correlation function for the
fluctuating field with a correlation time $\tau_0$. Then the
spectral densities are
\begin{equation}
    k_{qq}(\omega) = \overline{H_q^2}
    \frac{\tau_0}{1+\omega^2\tau_0^2}.
\end{equation}
In a physical system where $T_1^0$ is extremely long, we
approximate $1/T_1^0$ as zero in evaluating the effect of the
error suppression Hamiltonian on $T_2$ decay time. With this
approximation, $1/T_1$ is also zero since $k_{yy}(\omega_0\pm
J)\simeq k_{yy}(\omega_0)$ for $J\ll\omega_0$. The $T_2$ decay
time is then enhanced by
\begin{equation}
    \frac{T_2}{T_2^0} = \frac{k_{zz}(0)}{k_{zz}(J)} =
    1+(J\tau_0)^2
\end{equation}
due to the error suppression Hamiltonian. In the case of trapped
ions, $J$ can be made on the order of 10kHz \cite{Porras2004} and
the dominant noise component is at 50Hz \cite{Schmidt-Kaler2003}.
Assuming that this noise component is the maximum frequency
component, we obtain an enhancement of $T_2$ that is on the order
of $10^4$ by the error suppression Hamiltonian.

%\subsection{Comparison with active error correction}
Now we compare the current method to the conventional error
correction scheme. We consider a phase error due to the $T_2$
process at the rate $\epsilon = 1-e^{-\Delta t/T_2^0}$, in which
we perform error detection and correction at every time interval
$\Delta t$, using the three-qubit phase-flip code
\cite{Nielsen2000}. After the error correction, the error rate is
reduced to $3\epsilon^2$, corresponding to a lengthened effective
decay time $T_{2}^{\rm eff}$ defined by
\begin{equation}
    3\epsilon^2 = 1-e^{-\Delta t/T_{2}^{\rm eff}}.
\end{equation}

The $T_2$ increases as $(J\tau_0)^2$ for $J\tau_0\gg 1$ by natural
error suppression, while $T_2^{\rm eff}$ does linearly with the
number of error corrections per $T_2^{0}$ ($T_2^{0}/\Delta t$). As
summarized in Table~\ref{table:comparison}, the natural error
suppression method only requires simulating the static coupling
between qubits without the need for measurements or logic
operations, when the coupling can be made large enough. In
contrast, the conventional error correcting method requires a
large number of error detections and corrections, which involve
measurements and logic operations, to achieve the same enhanced
$T_2$ decay time.

\begin{table}[t]
    \centering
\begin{tabular}{c|c|c}
  \hline
  {Enhanced $T_2$} & {Natural error suppression} & {Active error
  correction} \\
  % after \\: \hline or \cline{col1-col2} \cline{col3-col4} ...
  $T_2/T_2^0$ & $J\tau_0$ & $T_2^0/\Delta t$ \\
  \hline
  $10^2$ & 10 & $3\times 10^2$ \\
  $10^4$ & $10^2$ & $3\times 10^4$  \\
  $10^6$ & $10^3$ & $3\times 10^6$  \\
   \hline
\end{tabular}
\caption{The enhanced $T_2$ decay time by natural error
suppression as a function of the coupling strength $J$ between two
qubits and the correlation time $\tau_0$ of the spectral densities
of fluctuating fields at the qubits. $T_2^0/\Delta t$ in the
rightmost column indicates the number of times per decay time
$T_2^0$ of error correction required to achieve the same enhanced
$T_2$. } \label{table:comparison}
\end{table}

%\subsection{Encoding and decoding}
Thus far, we have discussed the enhancement of $T_2$ by natural
error suppression for a stored qubit. Our storage method uses the
basis state $\ket{0}_{L}=\ket{++}$ and $\ket{1}_{L}=\ket{--}$ for
encoding. Suppose we have qubit information in the
$\ket{0}/\ket{1}$ basis, $c_0\ket{0}+c_1\ket{1}$, and encode it in
our storage. To do so, we first transfer the information to qubit
1 and apply the Hadamard transformation. Qubit 2 is initialized as
$\ket{+}$. A controlled-NOT gate between qubits 1 and 2,
conditioned on qubit 1, will lead the two-qubit state in
$c_0\ket{0}_{L} + c_1\ket{1}_{L}$. The decoding process is exactly
the reverse of the encoding process; we first apply a
controlled-NOT gate, conditioned on qubit 1, and then apply the
Hadamard transformation on qubit 1. Then the information is stored
in qubit 1 as $c_0\ket{0}+c_1\ket{1}$.

%\section{Discussion and conclusion}
We have shown that the decay time $T_2$ can be naturally
lengthened by simulating a coupling between physical qubits that
constitute an encoded qubit.  We discussed the case where the main
source for the $T_2$ process is fluctuation of precession
frequencies of the qubits in the transverse direction with no
energy dissipation (described by the term
$\gamma\sum_iH_{iz}I_{iz}$), driven by the zero-frequency
fluctuating field. Once the error suppression Hamiltonian is
introduced, the fluctuation of precession frequencies requires a
finite energy, driven by the higher-frequency fluctuation field.
When such an energy change is larger than the bandwidth of the
spectral densities of the fluctuating field, the fluctuation of
the precession frequencies are not allowed and therefore the $T_2$
process is suppressed. The condition for natural error suppression
to be effective is determined by comparison between the coupling
strength between physical qubits and the cut-off frequency of the
ambient fluctuating field. For slowly fluctuating field, even a
small coupling between physical qubits is effective in reducing
decoherence. Furthermore, unlike the DFS method, there is no
required symmetry in the qubit-reservoir coupling.

In this paper, we have shown the benefits of natural error
suppression by the simplest possible example -- a quantum memory
using the two-qubit phase error detecting code. Our proposed
method can easily extend to a quantum memory using three physical
qubits, whose ground state is the code subspace of the three-qubit
phase-flip error correcting code $\{\ket{+++}, \ket{---}\}$. The
stabilizer of the code consists of $\sigma_{1x}\sigma_{2x}$ and
$\sigma_{2x}\sigma_{3x}$ and the error suppression Hamiltonian can
be constructed solely with two-body interactions,
$-J(\sigma_{1x}\sigma_{2x}+\sigma_{2x}\sigma_{3x})$. With this
addition of an extra physical qubit, the conventional error
correction, with the help of ancilla qubits \cite{Nielsen2000} or
robust probe modes \cite{Yamaguchi2005}, can be applied to further
decrease the error rate when natural error suppression takes
place, or we can enjoy the automatic error correcting property
even when errors occur \cite{Barnes2000}. The method can also be
extended to suppress not only storage errors but also gate errors
for universal quantum computation, using an error-detecting code
involving more physical quits \cite{Bacon2001}. This will provide
an alternative approach to the DFS method, and alleviate the
overhead of conventional quantum error correction and
fault-tolerant quantum computation.

%In a case where a thermal reservoir is responsible for decay of
%physical qubits, fluctuation of precession frequencies of the
%qubits in the transverse direction necessarily involves an energy
%change $\omega_0$, since absorption or emission of a boson that
%constitutes the reservoir flips a qubit spin (described by a term
%$\propto b^{\dagger}I_{i-}+$ h.c., where $b^{\dagger}$ is the
%creation operator for the boson and $I_{i-}$ is the lowering
%operator of the qubit spin). With the error suppression
%Hamiltonian, the energy change $\omega_0$ is replaced by
%$\omega_0\pm J$. When $\omega_0\gg k_{\rm B}T$ and $\omega_0\gg J$
%as in the case with the trapped ion, the suggested natural
%suppression scheme has no significant effect on the decay process
%due to a thermal reservoir.

\noindent {\em Acknowledgments}: This work was supported in part
by JST SORST and NTT Basic Research Laboratories.

\bibliography{error_suppression}
\end{document}